%% file: halos.tex
\documentclass[twocolumn]{aastex63}

\received{March 9, 2021}
\revised{July 27, 2021}
\accepted{August 14, 2021}

\shorttitle{Effects of baryons on halos}
\shortauthors{Beltz-Mohrmann et al.}

\begin{document}

\title{The impact of baryonic physics on the abundance, clustering, and concentration of halos}

\correspondingauthor{Gillian D. Beltz-Mohrmann}
\email{gillian.d.beltz-mohrmann@vanderbilt.edu}

\author[0000-0002-4392-8920]{Gillian D. Beltz-Mohrmann}
\affiliation{Department of Physics and Astronomy \\ Vanderbilt University \\
Nashville, TN 37235, USA}

\author[0000-0002-1814-2002]{Andreas A. Berlind}
\affiliation{Department of Physics and Astronomy \\ Vanderbilt University \\
Nashville, TN 37235, USA}



\begin{abstract}
We examine the impact of baryonic physics on the halo distribution in hydrodynamic simulations compared to that in dark matter only (DMO) simulations. We find that, in general, DMO simulations produce halo mass functions (HMF) that are shifted to higher halo masses than their hydrodynamic counterparts, due to the lack of baryonic physics. However, the exact nature of this mass shift is a complex function of mass, halo definition, redshift, and larger-scale environment, and it depends on the specifics of the baryonic physics implemented in the simulation. We present fitting formulae for the corrections one would need to apply to each DMO halo catalogue in order to reproduce the HMF found in its hydrodynamic counterpart. Additionally, we explore the dependence on environment of this HMF discrepancy, and find that, in most cases, halos in low density environments are slightly more impacted by baryonic physics than halos in high density environments. We thus also provide environment-dependent mass correction formulae that can reproduce the conditional, as well as global, HMF. We show that our mass corrections also repair the large-scale clustering of halos, though the environment-dependent corrections are required to achieve an accuracy better than 2\%. Finally, we examine the impact of baryonic physics on the halo mass - concentration relation, and find that its slope in hydrodynamic simulations is consistent with that in DMO simulations. Ultimately, we recommend that any future work relying on DMO halo catalogues incorporate our mass corrections to test the robustness of their results to baryonic effects.
\end{abstract}



\section{Introduction} \label{sec:intro}
Studying the connection between galaxies and their dark matter halos is one of the keys to understanding galaxy formation and evolution, as well as constraining cosmological models. In recent years, using hydrodynamic simulations has become a popular method for investigating this connection \citep[e.g.][]{Vogelsberger2014}. However, these simulations are not only computationally expensive, but are also inconsistent. We currently lack a consensus on the correct baryonic physics prescriptions to use; thus, different hydrodynamic simulations produce widely varying results.

By contrast, dark matter only (DMO) simulations are much less computationally expensive, and although the only physics involved is gravity, they still allow us to predict the large-scale distribution of matter in the universe. However, a DMO simulation produces a halo mass function (HMF) that on average is shifted to higher masses than the HMF produced by a hydrodynamic simulation with the same cosmology and initial conditions \citep[e.g.,][]{Cui2014}. This is because in a hydrodynamic simulation the presence of baryonic physics like stellar feedback, star formation, and feedback from active galactic nuclei (AGN) has an impact on the masses of dark matter halos. This effect is non-trivial: the impact of baryonic physics varies with halo mass, as well as halo definition, environment, and redshift. And of course, the effect of baryonic physics depends on the details of the feedback prescriptions implemented, which varies from one hydrodynamic simulation to the next.

Several recent works have focused on comparing hydrodynamic simulations to models of the galaxy-halo connection \citep[e.g.][]{Hadzhiyska2020, Hadzhiyska2021, Hadzhiyska2021b}. \citet{Beltz-Mohrmann2020} examined the accuracy of halo occupation distribution (HOD) modeling compared to hydrodynamic simulations. They extracted HOD models from the galaxies in the Illustris and EAGLE simulations, and subsequently applied these HOD models to the corresponding DMO simulations (i.e. Illustris-Dark and EAGLE-Dark). They found that these HOD models, when applied to the DMO halos, were unable to reproduce the galaxy clustering seen in the hydrodynamic simulations. This was in part because the HMFs in the DMO simulations were shifted to higher masses, leading to an overestimate in the overall number density of galaxies, as well as discrepancies in other clustering statistics (e.g. correlation function and void probability function). The authors found that applying a correction to the DMO halo masses led to a significant improvement in the galaxy clustering. 

This issue with DMO halo mass functions could present a challenge for any work attempting to use halo modeling to constrain the galaxy-halo connection \citep[e.g.][]{Sinha2018}, as well as any work attempting to constrain cosmology \citep[e.g.][]{Lange2019}. If one's goal is merely to produce mock galaxy catalogues with the same clustering as seen in survey data, then one does not need to worry about this mass discrepancy; in this case, the parameterization of the galaxy-halo connection is unimportant, so long as the clustering mimics the real universe. However, if one's goal is to \textit{accurately} constrain the galaxy-halo connection, then this mass discrepancy presents a problem. Moreover, if one is attempting to use DMO simulations plus halo modeling to constrain cosmology, then obtaining the correct halo mass function is essential, unless the HOD is able to fully absorb the effect without perturbing the resulting cosmological parameters.

Several previous works have investigated the effects of baryonic physics on the halo mass function. \citet{Cui2012} compared three simulations: a dark matter only simulation, a hydrodynamic simulation with non-radiative physics, and a hydrodynamic simulation with radiative processes. They found that the fractional difference between halo masses in the hydrodynamic and DMO simulations is almost constant for halos above log$(M_{\Delta_c}/h^{-1}M_\odot) > 13.5$, but that for higher overdensity halos, as well as smaller mass halos, differences in halo mass appear which depend on halo mass as well as baryonic physics.

Later, \citet{Cui2014} further examined the effects of baryonic physics on the halo mass function, focusing on the role of AGN feedback. They found that for both friends-of-friends (FoF) and spherical overdensity (SO) halos, AGN feedback suppresses the HMF to a level below that of DMO simulations. They found that the ratio between the HMFs in the hydrodynamic and DMO simulations increases with overdensity, but does not have any redshift or mass dependence. They found that halos in hydrodynamic simulations have shallower inner density profiles, which lends them to halo mass loss caused by ``the sudden displacement of gas induced by thermal AGN feedback." The authors provide fitting functions to describe the halo mass variations between the full-physics and DMO simulations at different overdensities, which can recover the HMFs from hydrodynamic simulations for halo masses larger than $10^{13} h^{-1}M_\odot$. 

\citet{Sawala2013} examined the effect of baryons on the abundance of structures and substructures and found that halo masses are reduced for halos smaller than $10^{12}M_\odot$, and the effect grows with decreasing mass. Later, \citet{vanDaalen2014} looked at the impact of baryonic processes on the two-point correlation functions of galaxies, subhalos, and matter in large hydrodynamic simulations, and found that the changes due to the inclusion of baryons are not limited to small scales. The authors found that the large-scale effects are due to the change in subhalo mass caused by feedback associated with galaxy formation. They concluded that predictions of galaxy-galaxy and galaxy-mass clustering from models based on collisionless simulations will have errors greater than 10\% on scales $<1$ Mpc, unless the simulation results are modified to account for the effects of baryons on the distributions of mass and satellites. 

\citet{Velliscig2014} used hydrodynamic simulations from the OverWhelmingly Large Simulations (OWLS) project to study how the physical processes related to galaxy formation (e.g. star formation, supernova and AGN feedback, etc.) impact the properties of halos. They found that the ``gas expulsion and associated dark matter expansion induced by supernova-driven winds are important for halos with masses $M_{200} < 10^{13} M_\odot$, lowering their masses by up to 20\% relative a DM-only model." They also found that AGN feedback impacts halo masses up to cluster scales ($M_{200} \sim 10^{15} M_\odot$). Moreover, they found that baryonic physics alters the total mass profiles of halos out to several times the virial radius, which cannot be explained by a change in halo concentration. They concluded that the decrease in total halo mass leads to a decrease in the HMF of ~20\%. The authors provided analytic fitting formulae to correct halo masses and mass functions from DMO simulations. 

Subsequently, \citet{Khandai2015} investigated the properties of halos in the MassiveBlack-II hydrodynamical simulation, and found that baryons strongly affect the halo mass function when compared to dark-matter-only simulations, while \citet{Schaller2015} examined the effects of baryons on halos in the EAGLE simulation at low masses, and \citet{Chaves-Montero2016} used subhalo abundance matching to investigate the effect of baryons on the halo occupation distribution and assembly bias in the EAGLE simulation.

\citet{Bocquet2016} used the Magneticum simulations to investigate the impact of baryons on the halo mass function, and found that baryonic effects globally decrease the masses of galaxy clusters, which, at a given mass, results in a decrease of their number density. They found that this effect disappears at high redshift ($z\sim2$) and for high halo masses ($>10^{14} h^{-1} M_\odot$). They concluded that when using a survey like eROSITA, ignoring the impact of baryonic physics on the halo mass function leads to an underestimate in $\Omega_m$ by about 0.01. The authors also provided HMF fitting formulae.

\citet{Despali2017} examined the impact of baryonic physics on the subhalo populations in EAGLE and Illustris, and found that the presence of baryons reduces the number of subhalos, especially at the low-mass end. They found that the variations in the subhalo mass function depend on those in the halo mass function, which is shifted by the effect of stellar and AGN feedback.

Finally, \citet{Castro2021} (and earlier \citet{Balaguera2013}) addressed the impact of baryonic physics on clusters. They found that ignoring baryonic effects on the halos mass function and halo bias could significantly alter cosmological
parameter constraints, particularly in the upcoming generations of galaxy cluster surveys.

Most of these works concur that baryonic physics lead to a net reduction in the masses of halos, and consequently the HMF. However, the magnitude and mass dependence of the effect differ between different works, likely because of the different hydrodynamic simulations used. This calls for a systematic study that compares the impact of baryons on the HMF across a variety of recent hydrodynamic simulations with different baryonic physics and feedback prescriptions, and over a wide range of redshifts and halo definitions. Furthermore, no previous work has examined the environmental dependence of baryonic effects on the halo mass function, nor simultaneously examined the impact of baryonic physics on halo clustering and halo concentrations as a function of mass, all of which are important ingredients for halo models of galaxy clustering.

In this work, we investigate the impact of baryonic physics on the halo mass function in three different simulations: Illustris, IllustrisTNG, and EAGLE. We consider three different redshifts and five different halo definitions. Moreover, we investigate the environmental dependence of the effect of baryons on the halo mass function. For all these cases, we provide formulae that can be used to correct the halo masses from DMO simulations to match the halo mass functions from their hydrodynamic counterparts. Finally, we investigate the effect of baryonic physics on halo clustering and on the halo mass - concentration relation.

We discuss the simulation data in Section~\ref{sec:sims}. In Section~\ref{sec:hmf} we discuss the effects of baryons on halo populations, and in Section~\ref{sec:fits} we discuss our corrections. In Section~\ref{sec:env} we discuss the environmental dependence of the halo mass function, and in Section~\ref{sec:concentration} we discuss the halo mass - concentration relation. Finally, in Section~\ref{sec:conclusions} we summarize our results and conclusions.

\section{Simulation Data} \label{sec:sims}
For this analysis we use three cosmological N-body simulations: Illustris \citep{IllustrisPublicDataRelease,IntroducingIllustris,Vogelsberger2014,Genel2014}, IllustrisTNG \citep{Marinacci2018,Naiman2018,Nelson2018,Pillepich2018a,Pillepich2018b,Springel2018}, and EAGLE \citep{Schaye2015,McAlpine2016,EAGLE2017,Springel2005,Crain2015}. Specifically, we use Illustris-1, TNG100-1, TNG300-1, and EAGLE RefL100N1504. Each of these hydrodynamic simulations has a corresponding dark matter only counterpart. A summary of the simulation parameters can be found in Table~\ref{tab:simulations}.

The Illustris and IllustrisTNG simulations were performed with the \textsc{arepo} code, while the EAGLE simulation was performed with \textsc{gadget-3}. While all three simulations model star formation, stellar evolution, gas cooling and heating, supernovae feedback, black hole formation, and AGN feedback, \textsc{gadget-3} and \textsc{AREPO} contain different numerical hydrodynamical techniques, leading to large discrepancies in the galaxy populations they produce \citep{Scannapieco2012}. Additionally, though Illustris and TNG were performed with the same code, they vary in the strength of their feedback prescriptions. Illustris contains much stronger AGN feedback than either the TNG or EAGLE simulations \citep{Weinberger2017}, which leads to substantial differences in their results.

\input{simulations_table.tex}

We choose to compare these three simulations because they are publicly available and all have dark matter only (DMO) counterparts to their hydrodynamic simulations. Moreover, each simulation has adequate resolution and volume to allow us to examine its halo mass function between $10^{10}$ and $10^{14} h^{-1} M_\odot$. Finally, the three simulations have different baryonic physics prescriptions, which allows us to compare how variations in physics impacts our results.

In this work, we utilize halo catalogues from each simulation at $z=0,1,$ and $2$. In all three simulations, halos were identified using a standard friends-of-friends (FoF) algorithm \citep{Davis1985} with a linking length of $b=0.2$ times the mean interparticle separation. The FoF algorithm was run on the dark matter particles, and in the hydrodynamic simulations baryonic particles were attached to the same FoF group as their nearest dark matter particle. Each FoF group has a mass, which we refer to in this work as $M_\mathrm{FoF}$. 

Additionally, for each FoF group, masses were calculated for several different spherical overdensity (SO) definitions, which we will make use of in this paper. Specifically, these halo definitions are $M_{200b}$ (the total mass of the group enclosed in a sphere whose mean density is 200 times the mean density of the Universe at the time the halo is considered), $M_{200c}$ and $M_{500c}$ (the total mass of the group enclosed in a sphere whose mean density is 200 or 500 times the critical density of the Universe at the time the halo is considered), and $M_\mathrm{vir}$ (the total mass of this group enclosed in a sphere whose mean density is $\Delta_c$ times the critical density of the Universe at the time the halo is considered, where $\Delta_c$ derives from the solution of the collapse of a spherical top-hat perturbation from \citealt{Bryan1998}.) 

For each halo definition and redshift, we only consider halos above $10^{10} h^{-1} M_\odot$. In Illustris, the number of halos above this threshold ranges from about 60 to 101 thousand, while in IllustrisTNG, the number ranges from about 70 to 120 thousand, and in EAGLE, the number ranges from about 53 to 85 thousand. The exact number of halos depends on the halo definition and redshift, but it generally is lowest for $M_{500c}$ halos at $z=2$ and is highest for $M_\mathrm{FoF}$ halos at $z=1$.

\section{The Effect of Baryons on the Halo Mass Function} \label{sec:hmf}

\begin{figure}
	\includegraphics[width=\columnwidth]{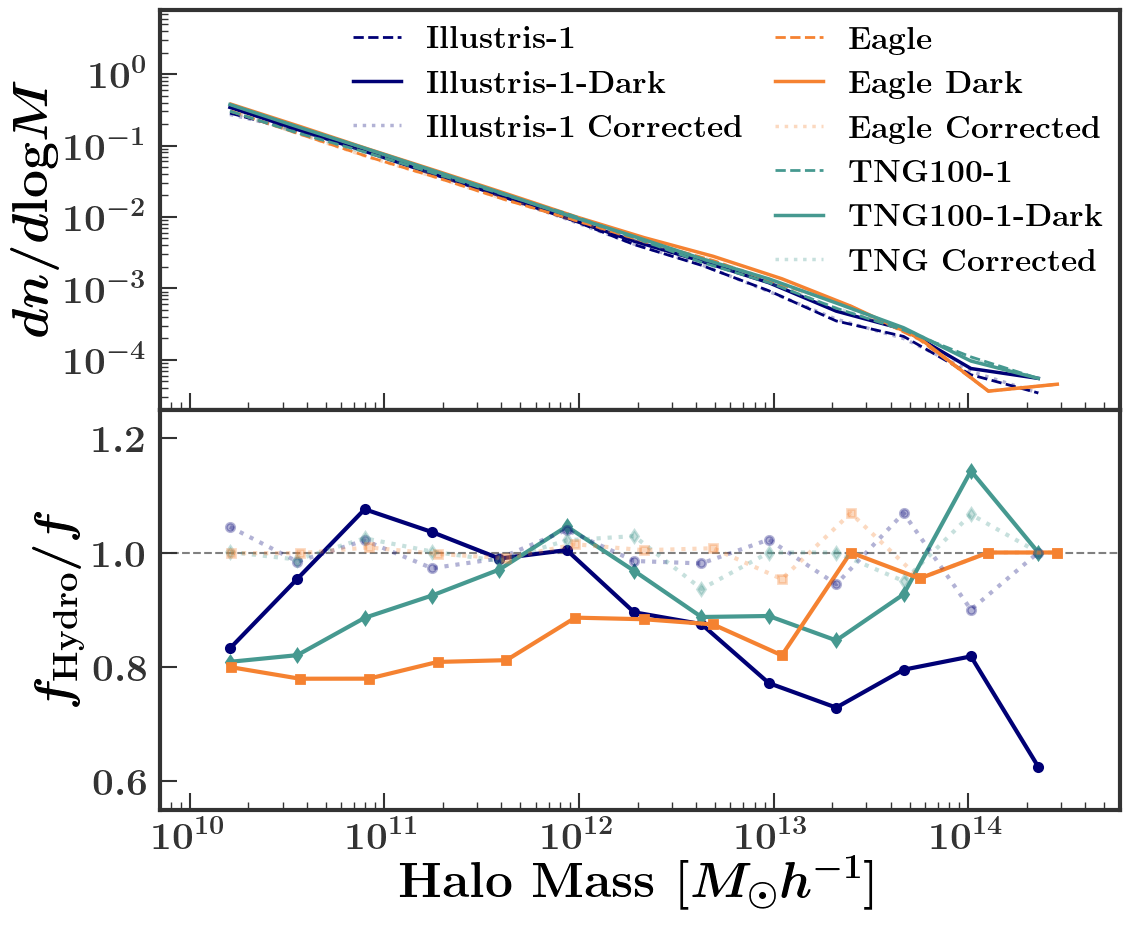}
    \caption{Halo mass functions of hydrodynamic compared to dark matter only simulations in the case of Illustris-1 (blue), TNG100-1 (green), and EAGLE RefL100N1504 (orange). The hydrodynamic versions are plotted with dashed lines, while the DMO versions are plotted with solid lines, and the corrected DMO versions are plotted with dotted lines. The bottom panel shows the ratio of the hydrodynamic HMF to either the DMO or the corrected DMO HMF for all three simulations. The halo definition shown is $M_{200b}$.}
    \label{fig:illustris_tng_eagle_hmf}
\end{figure}

The most straightforward way to investigate the population of halos in a simulation is by looking at the halo mass function (HMF), which displays the abundance of halos as a function of mass. In Figure~\ref{fig:illustris_tng_eagle_hmf}, we show the halo mass functions for $M_{200b}$ halos at $z=0$ for both the hydrodynamic and dark matter only versions of Illustris-1, TNG100-1, and EAGLE. Illustris is plotted in blue, TNG in green, and EAGLE in orange. It should be noted that the halo masses from the hydrodynamic simulations include both dark matter particles and baryonic particles. The hydrodynamic versions are plotted with dashed lines, while the DMO versions are plotted with solid lines. A corrected version of the DMO HMF is plotted with a dotted line and will be discussed further in the next section. In the residual panel, we plot the ratio of the hydrodynamic HMF to the DMO HMF (solid), as well as the ratio of the hydrodynamic HMF to the corrected DMO HMF (dotted) for each simulation.

The residuals reveal sizeable discrepancies between the DMO and hydrodynamic halo mass functions. In general, the hydrodynamic HMFs are shifted to lower masses in all three simulations, but this shift is mass dependent, and varies by simulation. In particular, the Illustris (blue) hydrodynamic HMF is consistently lower than the Illustris-Dark HMF above $10^{12} h^{-1} M_\odot$, as well as below $10^{11} h^{-1} M_\odot$. In TNG100 (green), the hydrodynamic HMF is lower than the DMO HMF between $10^{12}$ and $10^{14} h^{-1} M_\odot$, as well as below $10^{12} h^{-1} M_\odot$. In EAGLE (orange), the hydrodynamic HMF is lower than the DMO HMF at all halo masses below about $10^{13} h^{-1} M_\odot$.

In all three simulations, the discrepancies between the hydrodynamic and DMO halo mass functions reach the twenty percent level for much of the halo mass range. We emphasize that not only is this a significant difference, but it is also not trivial to correct. It is not equivalent to a difference in halo definition, which would be a constant offset and would not vary by simulation; this effect exhibits a trend with halo mass, and depends on the feedback implemented in the hydrodynamic simulation.

Our HMF results for Illustris are consistent with the findings of \citet{IntroducingIllustris}, who found that the halo mass function in Illustris is most affected at halo masses below $10^{10} h^{-1} M_\odot$ and above $10^{12} h^{-1} M_\odot$, where baryonic feedback processes (e.g. reionization and SN/AGN feedback) are strongest, leading to a reduction in halo mass compared to their DMO counterparts. They also found that halos around $10^{11} h^{-1} M_\odot$, where star formation is most efficient, tend to be more massive than their DMO counterparts.

Our HMF results for TNG100 are consistent with the findings of \citet{Springel2018}, who found that baryons in the TNG simulation have a larger impact on low mass halos and a smaller impact on high mass halos compared to Illustris. TNG has weaker AGN feedback than the original Illustris simulation, which leads to there being less discrepancy between the DMO and hydrodynamic HMFs at the high mass end. TNG also has a different wind model than Illustris, which leads to stronger feedback effects on lower mass halos, leading to more discrepancy at the lower mass end of the HMF in TNG.

Our HMF results for the EAGLE simulation are consistent with \citet{Desmond2017} and \citet{Schaller2015}, who examined the differences between the halo masses in the EAGLE DMO and hydrodynamic runs, and found the halos to be less massive on average in the hydrodynamic run. \citet{Desmond2017} found that, at low halo masses, stellar feedback in EAGLE strips baryons from the halo, which reduces the growth rate of the halo. At higher halo masses, stellar feedback becomes less effective, but AGN feedback is still capable of expelling baryons in all but the most massive halos.

\begin{figure*}
\centering
	\includegraphics[width=6.5in]{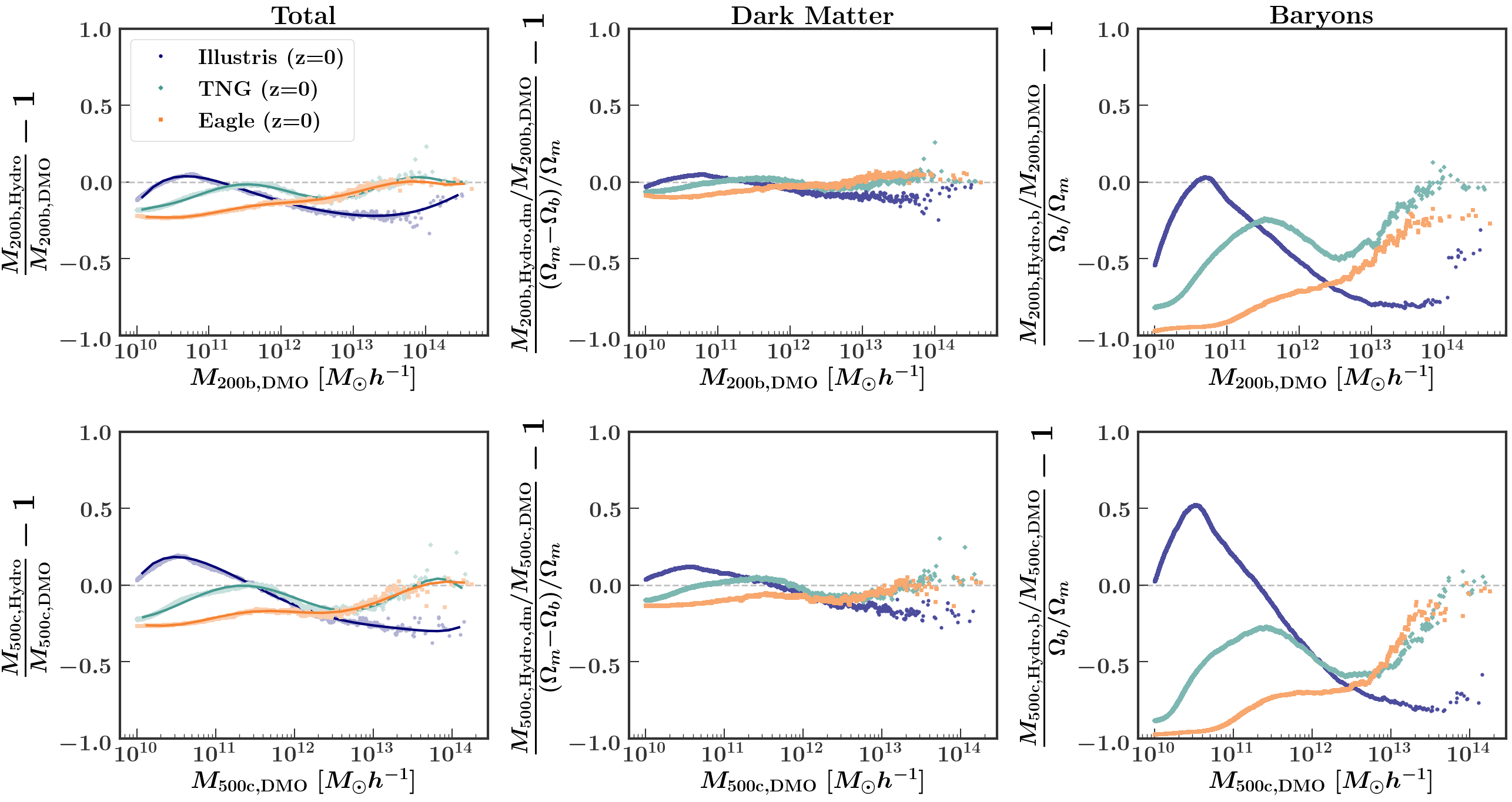}
    \caption{The fractional difference between halo masses from the hydrodynamic simulations to halo masses from the DMO simulations, as a function of DMO halo mass, all at $z=0$. The hydrodynamic mass in the y-axis of each column is (from left to right): total mass, mass of dark matter particles, and mass of baryonic (gas and star) particles. In the latter two cases, masses are normalized to account for the global difference between the total, dark matter, and baryon matter densities. The top row displays $M_{200b}$ halos, and the bottom row displays $M_{500c}$ halos. Hydrodynamic and dark matter only halos are matched by their mass rank, rather than by position. The displayed ratio thus represents the correction factor needed to apply to the dark matter only halos in order to recover the hydrodynamic halo mass function. Polynomial fit correction functions, as described in Section~\ref{sec:fits}, are plotted in the first column of panels as solid lines.}
    \label{fig:halo_masses_z0}
\end{figure*}

\begin{figure*}
\centering
	\includegraphics[width=6.5in]{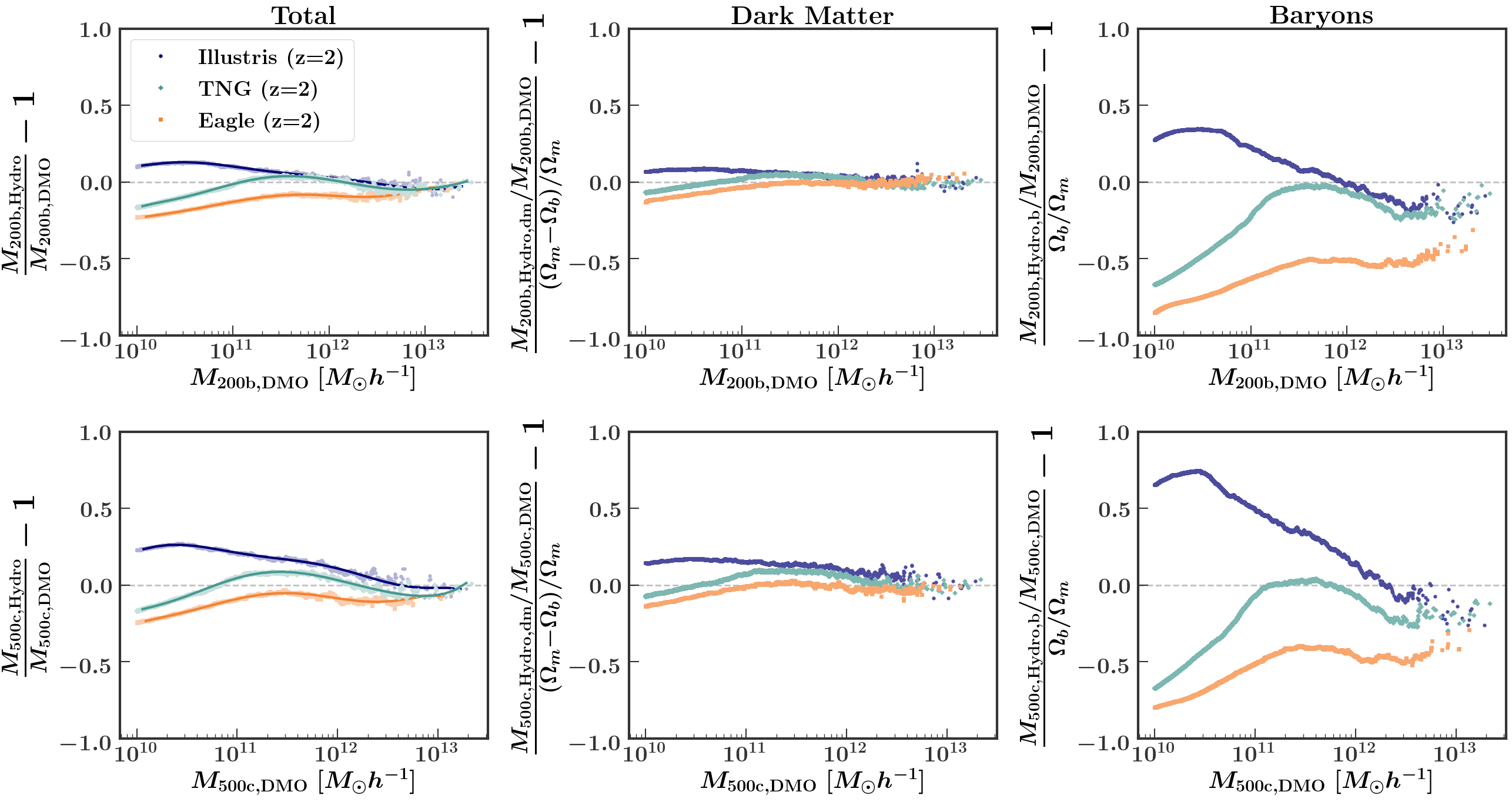}
    \caption{The fractional difference between halo masses from the hydrodynamic simulations to halo masses from the DMO simulations, as a function of DMO halo mass, all at $z=2$. All features are the same as in Fig.~\ref{fig:halo_masses_z0}.}
    \label{fig:halo_masses_z2}
\end{figure*}

Another way to visualize the discrepancy between the halo mass distributions from hydrodynamic and DMO simulations is to plot the fractional difference between their masses. In Figures~\ref{fig:halo_masses_z0} and \ref{fig:halo_masses_z2} we plot this fractional difference for $M_{200b}$ (top) and $M_{500c}$ (bottom) halos in Illustris, TNG100, and EAGLE at $z=0$ and $z=2$, respectively. Illustris halos are plotted in blue, TNG100 halos in green, and EALGE halos in orange. The x-axis is the DMO halo mass in units of $h^{-1} M_\odot$, and the y-axis is the fractional difference of hydrodynamic and DMO halo mass. In the left-hand column, we use the total hydrodynamic halo mass (i.e. dark matter and baryonic particles) to calculate this fractional difference. In the middle column, we use only the dark matter component of the hydrodynamic halo to calculate this fractional difference, and we normalize by one minus the universal baryonic mass fraction, $1 - \Omega_b/\Omega_m$, or $(\Omega_m - \Omega_b)/\Omega_m$. In the right-hand column, we use only the baryonic component (i.e. gas and star particles) of the hydrodynamic halo to calculate this fractional difference, and we normalize by the universal baryonic mass fraction $\Omega_b/\Omega_m$.

In each panel of Figures~\ref{fig:halo_masses_z0} and \ref{fig:halo_masses_z2} hydrodynamic and DMO halos are paired based on their ranked masses, rather than spatial positions or particle IDs. Thus, the most massive DMO halo is paired with the most massive hydrodynamic halo, and so on. In other words, we essentially ``abundance match" the halos in the hydrodynamic and DMO simulations (we do this separately for each redshift and halo definition, as well as each column in the figures; therefore, one cannot compare an individual point from one column or row to the next). We adopt this procedure because the fractional mass differences calculated in this way represent the correction one would need to apply to the halo masses from one of the DMO simulations in order to exactly recover the global halo mass function from the corresponding hydrodynamic simulation. While this abundance matching technique does not produce the exact same results as position or particle matching would, the overall trends seen are the same, although the scatter is drastically reduced by abundance matching \citep{IntroducingIllustris,Schaller2015,Springel2018}.

Examining the top left panel of Figure~\ref{fig:halo_masses_z0}, we can see that the same trends are present here as were present in the halo mass functions shown in Figure~\ref{fig:illustris_tng_eagle_hmf}. The results at $z=0$ for $M_{200b}$ halos in Illustris (consistent with \citealt{IntroducingIllustris} and \citealt{Springel2018}) indicate that stellar feedback slightly reduces the masses of the lowest mass halos (by up to $\sim10\%$), while star formation efficiency slightly increases the masses of halos around $10^{11} h^{-1} M_\odot$, and AGN feedback severely reduces the masses of high mass halos (by up to $\sim20\%$). The results for TNG100 (consistent with \citealt{Springel2018}) indicate that stellar feedback reduces the masses of low mass halos (by up to $\sim15\%$), while star formation efficiency peaks at slightly higher masses than in Illustris (but is not quite so efficient as to increase halo masses), while AGN feedback is less strong than in Illustris, and does not effect the highest mass halos, but reduces the masses of intermediate mass halos by up to $\sim10\%$. The results for EAGLE (consistent with \citealt{Schaller2015}) indicate that stellar feedback severely reduces the masses of low mass halos (by up to $\sim20\%$), while AGN feedback is similar to that in TNG, and reduces the masses of intermediate mass halos (by $\sim10\%$) but does not impact the highest mass halos.

Looking at the second and third columns of Figure~\ref{fig:halo_masses_z0}, we can see that feedback has a much more extreme effect on the baryons in a halo than it does on the dark matter particles. Thus, most of the mass difference is due to a loss or gain of baryons, and not dark matter. For example, in the EAGLE simulation, there is an extreme lack of baryons at lower halo masses, while in the Illustris simulation, there is a significant lack of baryons at higher halo masses, but for both simulations, the amount of dark matter in each halo remains relatively constant.

The differences between the three simulations in the third column of Figure~\ref{fig:halo_masses_z0} indicate that these three hydrodynamic simulations disagree in terms of the baryon content of their halos. Precise observational constraints on the baryon fraction as a function of halo mass could in principle allow us to differentiate between these three simulations. \citet{Gonzalez2013} measured the baryons contained in both the stellar and hot-gas components for galaxy clusters and groups with $M_{500} > 10^{14} M_\odot$ at $z=0.1$, and found that the weighted mean baryon fraction for halos with $M_{500} > 2 \times 10^{14} M_\odot$ is 7\% below the universal value when using a Planck cosmology. This is consistent with the results from the TNG and EAGLE simulations at the high mass end, but is not in agreement with the results from the Illustris simulation. The more significant differences between the three simulations, however, occur at lower halo masses. \citet{Eckert2017} examined the baryonic content of halos in the ECO and RESOLVE galaxy surveys \citep{Moffett2015}. While these results do extend to lower halo masses, they only include cold baryonic content, so we cannot make a direct comparison to the baryonic content in the three hydrodynmaic simulations. In the future, more precise observational constraints on the baryon content of low-mass halos could allow us to rule out certain hydrodynamic simulations.

Comparing the top and bottom rows of Figure~\ref{fig:halo_masses_z0}, we can see that the discrepancy between hydrodynamic and DMO halo masses is more extreme for $M_{500c}$ halos than it is for $M_{200b}$ halos, indicating that feedback has stronger impact on the inner regions of a halo. This is most evident when looking at the baryonic component of Illustris $M_{500c}$ halos (in the third column of the second row of Figure~\ref{fig:halo_masses_z0}), where feedback and star formation have an extreme effect on the percentage of baryons in a halo.

We can also see that the trends present in each simulation at $z=0$ are quite different from those at $z=2$, which are shown in Figure~\ref{fig:halo_masses_z2}. For example, we can see clearly from the $M_{500c}$ Illustris halos that at $z=2$, stellar and AGN feedback have not kicked in yet, and star formation efficiency is quite strong, resulting in a strong overabundance of gas and star particles. Because of this, essentially all $M_{500c}$ Illustris halos at $z=2$ have a higher mass in the hydrodynamic simulation than they do in the DMO simulation (by $\sim10\%$). However, by $z=0$, stellar feedback has kicked in for halos at or below $10^{10} h^{-1} M_\odot$, while AGN feedback has kicked in for halos above $5 \times 10^{11} h^{-1} M_\odot$, resulting in these halos having lower masses in the hydrodynamic simulation than their DMO counterparts. Meanwhile, in TNG100 and EAGLE, it appears that stellar feedback has already kicked in by $z=2$, but AGN feedback has not, leading to less of a reduction in halo mass at the high mass end, but a similar result at low masses.

\begin{deluxetable*}{ccccccccccc}
\tablenum{2}
\tablecaption{Dark matter only halo mass corrections. \label{tab:fits0}}
\tablewidth{0pt}
\tablehead{
\colhead{Halo Def.} & \colhead{Sim.} & \colhead{Lim.}  & \colhead{a} & \colhead{b} & \colhead{c} & \colhead{d} & \colhead{e} & \colhead{f} & \colhead{g} & \colhead{h}
}
\startdata
{} & Illustris & $3.2 \times 10^{14}$ & -0.000215 & 0.024375 & -0.004406 & -0.05112 & 0.32264 & -0.75977 & 0.65421 & -0.14617 \\
$M_{200b}$ & TNG & $3.6 \times 10^{14}$ & 0.0035204 & -0.05293 & 0.30055 & -0.79135 & 0.94957 & -0.45741 & 0.17801 & -0.19176 \\
{} & EAGLE & $4.3 \times 10^{14}$ & 0.0019695 & -0.02991 & 0.17485 & -0.49145 & 0.66861 & -0.36278 & 0.06311 & -0.2312 \\
\hline
{} & Illustris & $3.5 \times 10^{14}$ & -0.0007237 & 0.010774 & -0.05997 & 0.13949 & -0.03321 & -0.3973 & 0.47075 & -0.10679 \\
$M_\mathrm{FoF}$ & TNG & $4.1 \times 10^{14}$ & 0.002596 & -0.038396 & 0.21334 & -0.54396 & 0.61565 & -0.26124 & 0.11294 & -0.18982 \\
{} & EAGLE & $4.3 \times 10^{14}$ & 0.001614 & -0.024094 & 0.13731 & -0.37018 & 0.46557 & -0.2016 & 0.007495 & -0.20007 \\
\hline
{} & Illustris & $2.8 \times 10^{14}$ & -0.0003252 & 0.003505 & -0.006643 & -0.059713 & 0.3689 & -0.083365 & 0.67904 & -0.12001 \\
$M_\mathrm{vir}$ & TNG & $3.2 \times 10^{14}$ & 0.003404 & -0.05052 & 0.28087 & -0.71152 & 0.78559 & -0.30528 & 0.13288 & -0.19697 \\
{} & EAGLE & $3.4 \times 10^{14}$ & 0.0021115 & -0.031718 & 0.18264 & -0.50231 & 0.66224 & -0.34231 & 0.057532 & -0.23903 \\
\hline
{} & Illustris & $2.0 \times 10^{14}$ & -0.0015556 & 0.019282 & -0.084937 & 0.12921 & 0.14682 & -0.73173 & 0.65319 & -0.069491 \\
$M_{200c}$ & TNG & $2.4 \times 10^{14}$ & 0.003539 & -0.050576 & 0.26672 & -0.61706 & 0.54882 & -0.073071 & 0.075193 & -0.20515 \\
{} & EAGLE & $2.5 \times 10^{14}$ & 0.0024825 & -0.036885 & 0.20893 & -0.55934 & 0.70466 & -0.33643 & 0.051158 & -0.24832 \\
\hline
{} & Illustris & $1.4 \times 10^{14}$ & 0.0016927 & -0.024244 & 0.14307 & -0.46591 & 0.96494 & -1.3004 & 0.76757 & 0.029836 \\
$M_{500c}$ & TNG & $1.5 \times 10^{14}$ & 0.0017185 & -0.026947 & 0.14529 & -0.30347 & 0.13573 & 0.14566 & 0.08892 & -0.21709 \\
{} & EAGLE & $1.7 \times 10^{14}$ & 0.002766 & -0.040727 & 0.22567 & -0.57591 & 0.65174 & -0.22731 & 0.01389 & -0.26156
\enddata
\tablecomments{Shown here are the DMO halo mass corrections for $z=0$ halos. The columns list (from left to right): halo definition, simulation, mass limit, and the seventh order polynomial coefficients of the halo mass correction fits, beginning with the highest order. Table 2 is published in its entirety (including corrections for halos at $z=1$ and $z=2$) in the machine-readable format.}
\end{deluxetable*}

\section{Correcting the DMO Halo Mass Function} \label{sec:fits}
Overall, Figures~\ref{fig:halo_masses_z0} and \ref{fig:halo_masses_z2} emphasize the fact that the effect of baryonic physics on the halo mass function is to shift the HMF to lower masses. This shift can be as large as 25 percent. However, it is clear that this effect varies dramatically from one simulation to the next. This presents a problem for anyone using DMO halo catalogues to constrain the galaxy-halo connection or cosmology \citep{Beltz-Mohrmann2020}. The solution to this problem is not as simple as adding a parameter to the model, because it is a problem with the dark halo population itself. Additionally, the solution is not as simple as applying a constant offset to all DMO halo masses, because the discrepancy depends on the mass regime. Finally, once again, the solution depends on which hydrodynamic simulation is regarded as having the correct baryonic physics.

One possible solution is to apply a correction to each of the halos in a DMO simulation, so that the HMF better mimics the mass function from a hydrodynamic simulation. This correction should serve to adjust each DMO halo so that it has the mass of its corresponding hydrodynamic halo. We can use the fractional difference in halo mass shown in Figures~\ref{fig:halo_masses_z0} and \ref{fig:halo_masses_z2} to identify a functional form for this correction. 

To do this, we fit a polynomial to the fractional difference in halo masses between our hydrodynamic and DMO simulations (i.e., column one of Figures~\ref{fig:halo_masses_z0} and \ref{fig:halo_masses_z2}). The polynomial fits to each halo mass relationship were found using NumPy's polyfit function. After examining the effectiveness of several polynomials to correct the halo mass function, we found that a 7th order polynomial was the lowest order that could accurately capture the halo mass trend and allow us to correct the DMO halo mass function. We have additionally looked at the Bayesian Information Criteria (BIC) for polynomials ranging from 3rd order to 12th order and found that the mean BIC (averaged over each halo definition and redshift combination) continues to decrease until we reach 7th order, but does not continue to decrease significantly for higher orders. Thus, we decided that a 7th order polynomial was the lowest order we could use for our fits and still accurately correct the halo mass function.

The corrected DMO halo masses (in units of $10^{10} h^{-1} M_\odot$) are given by
\begin{equation}
M_\mathrm{h,corrected} = (y + 1) \times M_\mathrm{h,DMO}
\end{equation}
where $M_\mathrm{h,DMO}$ is the unlogged original halo mass in units of  $10^{10} h^{-1} M_\odot$, and
\begin{equation}
y = ax^7 + bx^6 + cx^5 + dx^4 + ex^3 + fx^2 + gx + h ,
\end{equation}
where $x = \mathrm{log}_{10}(M_\mathrm{h,DMO})$ and $a$ through $h$ are the polynomial coefficients for a given simulation and halo definition. These fits (for $M_{200b}$ and $M_{500c}$ halos at redshifts 0 and 2) are plotted with solid lines in the first column panels of Figures~\ref{fig:halo_masses_z0} and \ref{fig:halo_masses_z2}. The fits at redshift 0 for all halo definitions are listed in Table~\ref{tab:fits0}. Additionally, fits for these same five halo definitions at redshifts 1 and 2 are available in the machine-readable format.

Once again, this correction in based on ``abundance matching" the halos between hydrodynamic and DMO simulations. Thus, applying this correction will assign to the most massive DMO halo the mass of the most massive hydrodynamic halo, and so on. After applying our corrections to each DMO simulation, we can examine our new corrected DMO halo mass functions. In Figure~\ref{fig:illustris_tng_eagle_hmf}, we have plotted these corrected HMFs for $M_{200b}$ halos at $z=0$ with dotted lines. Looking at the residual panel, we can see that the corrections lead to significant improvement, and almost perfectly reproduce the HMFs from the hydrodynamic simulations (deviations are less than $5\%$), which was precisely the goal of our abundance matching method.

Unfortunately, the Illustris, TNG100, and EAGLE simulations do not contain any halos with masses above about $4 \times 10^{14} h^{-1} M_\odot$ at $z=0$. This upper limit decreases for $z=1$ and $z=2$ halos, and is also somewhat dependent on halo definition. We do not make any assumptions about the masses of our halos beyond our data. Thus, our mass corrections should not be applied to any DMO halos above the limits given in Table~\ref{tab:fits0}. Rather, any DMO halo above our limits should be left unaltered. (This is very important; because our mass corrections are seventh order polynomials, extending them beyond our mass limits will lead to very large changes in halo mass.)

It is noteworthy that in TNG100 and EAGLE, the mass correction is already almost zero at the high mass end. However, in Illustris, for $z=0$ halos, this is not the case. This means that applying the Illustris halo mass correction will lead to a slight discontinuity in the halo mass function. This discontinuity could be alleviated by, for example, extrapolating the fit at the high mass end; however, because doing this would not be based on any data, we do not provide any extrapolations of our fits here. Additionally, our mass corrections should only be applied to halos with masses above $10^{10} h^{-1} M_\odot$. We do not present corrections for halos below $10^{10} h^{-1} M_\odot$ in this work due to the mass resolutions of the simulations that we use. (Once again, it is very important not to apply the corrections to any DMO halos below $10^{10} h^{-1} M_\odot$ due to the nature of the seventh order polynomial fits.)

We recommend that any future work using halos from a DMO simulation do the following: for a given halo definition and redshift, apply at least one of our halo mass corrections to correct the halo mass function. Once the mass corrections are applied, the remaining analysis (e.g. HOD, CLF, etc.), can be performed on the corrected halo catalogues. In this way, one can investigate the robustness of their results to changes in the halo mass function. Ideally, we recommend applying all three corrections (i.e. each correction based on Illustris, TNG100, and EAGLE) since the variation among them represents the theoretical uncertainty in baryonic physics.

We have created a \textsc{python} module for implementing our halo mass corrections, which is publicly available at \url{https://github.com/gbeltzmo/halo_mass_correction}. This module takes in an array of halo masses, a halo definition ($M_{200b}$, $M_\mathrm{FoF}$, $M_\mathrm{vir}$, $M_{200c}$, or $M_{500c}$), a redshift (0, 1, or 2), and a simulation (Illustris, TNG, or EAGLE), and returns the corresponding corrected halo masses. If a given halo mass is outside the accepted mass range, the code will issue a warning, and will return the original (uncorrected) halo mass.

One question worth investigating is whether applying our mass corrections to a box with a very different resolution produces accurate results. To investigate this, we applied the TNG correction (based on the TNG100-1 simulation) to the TNG300-1-Dark box (which is about a factor of 8 lower in resolution than the TNG100-1-Dark simulation) to see if we could reproduce the halo mass function from the TNG300-1 simulation. Looking only at the $10^{10} - 10^{14} h^{-1} M_\odot$ mass range, this correction almost perfectly reproduces the halo mass function from TNG300-1, with all deviations less than 5\% (compared to 17\% deviations without the correction). This indicates that as long as the mass corrections are only applied within the appropriate mass regime, the corrections are accurate even when used with simulations of different volumes and resolutions.

Szewciw et al. (2021, in preparation) applied our mass corrections to $M_\mathrm{vir}$ halos (z=0) from a large DMO simulation \citep[Las Damas;][]{McBride2009} and examined how their HOD parameter constraints on SDSS galaxies varied with the different halo mass corrections. They found that for both their luminosity samples ($M_r^{-19}$ and $M_r^{-21}$), the different halo mass corrections lead to changes in all their HOD parameters. The biggest changes are seen in $\mathrm{log}{M_\mathrm{min}}$ and $\mathrm{log}{M_1}$, and in most cases the Illustris-based mass correction leads to the biggest change, although for the $M_r^{-19}$ sample it is the EAGLE-based correction that leads to the biggest difference in $\mathrm{log}{M_\mathrm{min}}$. While it is to be expected that the halo mass corrections lead to changes in the mass parameters of the HOD, these changes are simulation dependent, and are not trivial to predict. The halo mass corrections ultimately do not lead to better $\chi^2$ values for the best fitting models, nor do they lead to the models being ruled out.

\section{Environmental Dependence} \label{sec:env}
Our ``abundance matching" technique does not explicitly take halo environment into account when correcting the halo masses. This means that while our mass corrections successfully reproduce the global halo mass function from our hydrodynamic simulations, they will not correct the \textit{conditional} HMF if baryonic effects on halo mass are environment dependent. 

Fitting halo mass corrections in which halos are matched between DMO and hydro based on position or particle IDs would inherently take halo environment into account. However, there are several issues with this technique. Firstly, matching halos based on position or particle IDs does not guarantee that every halo in the hydrodynamic simulation has a match in the DMO simulation. Secondly, this method of matching introduces a significant amount of scatter into the hydrodynamic-to-DMO halo mass relationship. Therefore, if one were to use this relationship to correct halos from a large DMO simulation, one would have to either ignore the scatter (in which case the result is essentially the same as the abundance matching correction), or take the scatter into account by binning halos by mass and then drawing their corrected mass from a distribution within that bin. While this accounts for the scatter, it still does not account for halo environment. Additionally, this is not the cleanest method for correcting DMO halo masses; when applied to any of the three simulations used in this work, this method does not successfully reproduce the correct global \textit{or} conditional HMF from the hydrodynamic simulations.

One alternate possibility is to use our original ``abundance matching" technique, but to separate halos by environment. We can do this by measuring the large-scale environment around each of our DMO and hydrodynamic halos. We can then split our halos into those with high-density environments and those with low-density environments, and subsequently ``abundance match" between DMO and hydro, matching \textit{only} halos within \textit{similar} environments. For example, the most massive DMO halo in a \textit{high}-density environment would be matched with the most massive hydrodynamic halo in a \textit{high}-density environment, while the most massive DMO halo in a \textit{low}-density environment would be matched with the most massive hydrodynamic halo in a \textit{low}-density environment, and so on. This procedure will yield mass corrections that are guaranteed to recover the correct conditional HMF.

To do this for each of our simulations, we first have to measure the large-scale environment around each of our halos. To measure halo environment, we calculate the total mass of \textit{halos} within 5 Mpc spheres centered on each halo of interest (excluding the mass of the halo of interest itself). In other words, we do not sum up all particles, but rather sum up the masses of all halos in the halo catalog whose centers fall within the sphere (excluding the main halo). We do not impose any lower mass limit on the halos included in this sum. We measure environments for all DMO halos above $10^{10} h^{-1} M_\odot$, and all hydrodynamic halos that are matched to them using the abundance matching method described above.

We can thus define an environment factor $\delta$ for each halo, such that $\delta = (\rho_\mathrm{sphere} / \rho_\mathrm{box}) - 1$, where $\rho_\mathrm{sphere}$ is the mass of halos in a 5 Mpc sphere around the halo divided by the volume of a 5 Mpc sphere, and $\rho_{box}$ is the sum of all halo masses in the box divided by the volume of the box. We measure halo environment in this way for all three of our simulations - hydrodynamic and DMO alike. We repeat this measurement for each of our different halo definitions and redshifts. We then split our halos into ``high-" and ``low-density" environments based on the median environment ($\delta_\mathrm{med}$) for that simulation. This is done separately for each halo definition and redshift, and is also done separately for the hydrodynamic and DMO simulations. Thus, each simulation/halo definition/redshift combination has a slightly different $\delta_\mathrm{med}$. Subsequently, DMO halos in high-density environments are ``abundance matched" with hydrodynamic halos also in high-density environments, and likewise for halos in low-density environments. 

\begin{figure}
	\includegraphics[width=\columnwidth]{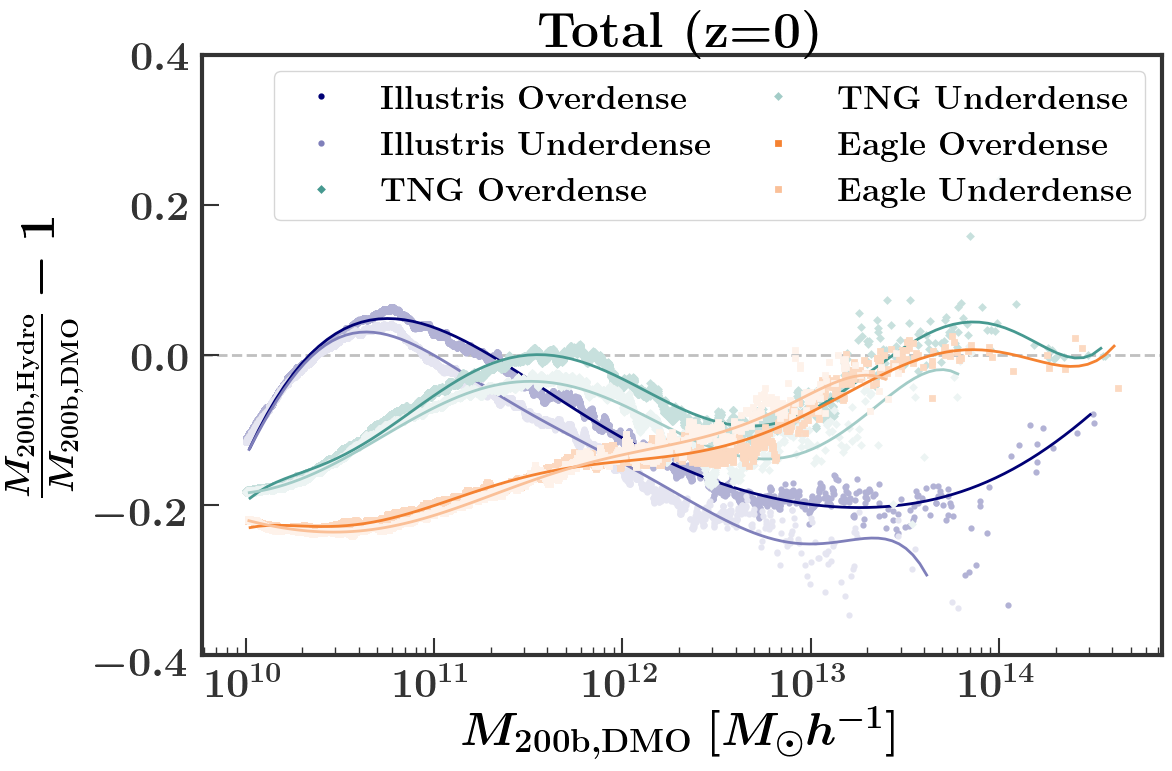}
    \caption{Fractional difference between hydrodynamic and DMO $M_{200b}$ halos from Illustris-1 (blue), TNG100-1 (green), and EAGLE (orange) at $z=0$. In this plot, hydrodynamic and DMO halos are first split into high and low density environments, and then are abundance matched with corresponding halos in similar environments. High density environments are plotted in darker colors, and low density environments are plotted in lighter colors. Each trend is fit with a seventh order polynomial, which is given in Table~\ref{tab:envfits0}.}
    \label{fig:env}
\end{figure}

Shown in Figure~\ref{fig:env} are the results of this environment-dependent abundance matching technique for $M_{200b}$ halos at $z=0$ in Illustris-1 (blue), TNG100-1 (green), and EAGLE (orange). Halos in high-density environments are plotted in darker colors, and halos in low-density environments are plotted in lighter colors. Each of these relationships is once again fit with a seventh order polynomial, which is plotted here with a solid line. The fits for $M_{200b}$ and $M_\mathrm{vir}$ halos at $z=0$, along with the $\delta_\mathrm{med}$ and mass limit, are given in Table~\ref{tab:envfits0}.

\begin{figure}
	\includegraphics[width=\columnwidth]{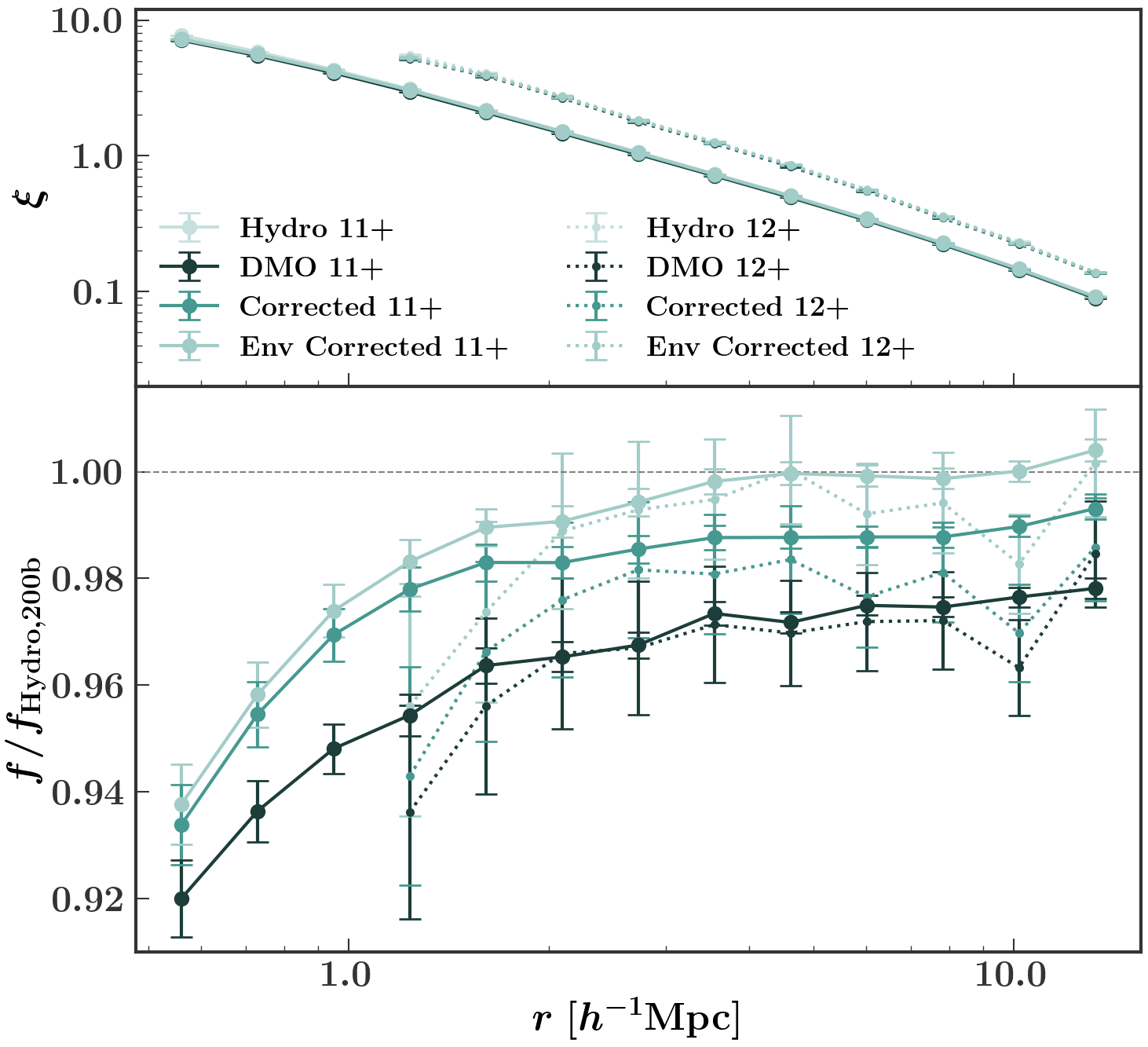}
    \caption{The halo correlation functions for $M_{200b}$ halos at $z=0$ in TNG300-1 (pale green), TNG300-1-Dark (dark green), TNG300-1-Dark with the abundance matching correction (green), and TNG300-1-Dark with the environment-dependent abundance matching correction (light green). Halo correlation functions are plotted for halos greater than $10^{11} h^{-1} M_\odot$ (solid), and greater than $10^{12} h^{-1} M_\odot$ (dotted). The bottom panel shows the ratios of each DMO halo correlation function to TNG300-1 for a given mass bin.}
    \label{fig:xi}
\end{figure}

We can see from Figure~\ref{fig:env} that for each simulation, the relationship between hydrodynamic and DMO halos closely resembles that seen in Figure~\ref{fig:halo_masses_z0} for $M_{200b}$ halos, but we do detect a difference between the high- and low-density environments. In Illustris and TNG100, the differences between environments appear for all DMO halos above $5 \times 10^{10} h^{-1} M_\odot$, where hydrodynamic halos in high-density environments are slightly more massive relative to DMO halos than those in low-density environments. In EAGLE, below $2 \times 10^{10} h^{-1} M_\odot$ the high- and low-density trends are the same; between $2 \times 10^{10}$ and $5 \times 10^{11} h^{-1} M_\odot$ the hydrodynamic halos in high-density environments are slightly more massive relative to DMO halos than those in low-density environments; at higher masses this relationship is reversed. Additionally, for all three simulations, the highest mass hydrodynamic and DMO halos are exclusively found in high-density environments, as expected.

While our original ``abundance matching" halo mass correction reproduces the global halo mass function from a given hydrodynamic simulation, our environment-dependent halo mass correction reproduces both the global \textit{and} the conditional halo mass function. To examine whether it is important to implement this more complicated halo mass correction, we can see how well our different mass corrections reproduce the halo clustering found in hydrodynamic simulations by measuring the halo correlation function, $\xi(r)$. 

\begin{deluxetable*}{ccccccccccccc}
\tablenum{3}
\tablecaption{Environment dependent halo mass corrections. \label{tab:envfits0}}
\tablewidth{0pt}
\tablehead{\colhead{Halo} & \colhead{Sim.} & \colhead{$\delta_{med}$} & \colhead{Env.} & \colhead{Lim.} & \colhead{a} & \colhead{b} & \colhead{c} & \colhead{d} & \colhead{e} & \colhead{f} & \colhead{g} & \colhead{h}}
\startdata
{} & I & 0.056 & High & $3.2 \times 10^{14}$ & -0.000282 & 0.00451 & -0.025829 & 0.050903 & 0.082441 & -0.49837 & 0.55978 & -0.13646 \\
{} & {} & {} & Low & $6.0 \times 10^{13}$ & -0.009371 & 0.11086 & -0.50638 & 1.0971 & -1.0281 & 0.016647 & 0.45108 & -0.13212 \\
\cline{2-13}
$M_{200b}$ & T & 0.03 & High & $3.6 \times 10^{14}$ & 0.004054 & -0.06133 & 0.3523 & -0.94844 & 1.19187 & -0.63492 & 0.23833 & -0.19604 \\
{} & {} & {} & Low & $7.1 \times 10^{13}$ & -0.000712 & 0.002265 & 0.018051 & -0.074849 & 0.015843 & 0.1256 & 0.027276 & -0.1839 \\
\cline{2-13}
{} & E & 0.064 & High & $4.3 \times 10^{14}$ & 0.001995 & -0.030087 & 0.17401 & -0.48138 & 0.64179 & -0.34464 & 0.069041 & -0.23138 \\
{} & {} & {} & Low & $2.9 \times 10^{13}$ & 0.000006 & -0.005095 & 0.050413 & -0.17942 & 0.26066 & -0.089445 & -0.027582 & -0.2228 \\
\hline
{} & I & 0.068 & High & $2.8 \times 10^{14}$ & 0.000327 & -0.006513 & 0.053503 & -0.23813 & 0.64263 & -1.05189 & 0.7843 & -0.13659 \\
{} & {} & {} & Low & $4.9 \times 10^{13}$ & -0.008 & 0.090915 & -0.39228 & 0.76673 & -0.50891 & -0.42139 & 0.60965 & -0.12534 \\
\cline{2-13}
$M_\mathrm{vir}$ & T & 0.039 & High & $3.2 \times 10^{14}$ & 0.003833 & -0.057633 & 0.32688 & -0.8577 & 1.021834 & -0.49292 & 0.20724 & -0.20621 \\
{} & {} & {} & Low & $6.4 \times 10^{13}$ & -0.009157 & 0.10013 & -0.42418 & 0.91773 & -1.1416 & 0.77501 & -0.098584 & -0.18654 \\
\cline{2-13}
{} & E & 0.073 & High & $3.4 \times 10^{14}$ & 0.002174 & -0.032846 & 0.19068 & -0.53199 & 0.72558 & -0.4227 & 0.10743 & -0.24488 \\
{} & {} & {} & Low & $2.5 \times 10^{13}$ & 0.001936 & -0.029656 & 0.17358 & -0.4863 & 0.65787 & -0.35221 & 0.059215 & -0.24129
\enddata
\tablecomments{Shown here are the environment dependent halo mass corrections for $M_{200b}$ and $M_\mathrm{vir}$ halos at $z=0$ in Illustris (I), TNG (T), and EAGLE (E). The first four columns are the halo definition, simulation, the median environment factor $\delta$, and the type of environment. The fifth column is the upper mass limit for each fit. (The lower limit is $10^{10} h^{-1} M_\odot$ for all fits.) The remaining columns are the polynomial coefficients for the fits. Table 3 is published in its entirety (including corrections for the remaining halo definitions and redshifts) in the machine-readable format.}
\end{deluxetable*}

In Figure~\ref{fig:xi} we plot the halo correlation functions for $M_{200b}$ halos at $z=0$ in TNG300-1. We use the TNG300-1 box for this analysis because the smaller hydrodynamic boxes contain too few halos to accurately examine the clustering of halos. We calculate the halo correlation function in two different mass bins: halos greater than $10^{11} h^{-1} M_\odot$ and halos greater than $10^{12} h^{-1} M_\odot$. For the lower mass sample, we measure $\xi$ in 13 bins of separation $r$ between $0.49$ and $15 h^{-1}\mathrm{Mpc}$, and for the higher mass sample we measure $\xi$ in 10 bins of separation $r$ between $1.07$ and $15 h^{-1}\mathrm{Mpc}$. We compute $\xi$ with the blazing fast code \textsc{corrfunc} \citep{corrfunc2017, Corrfunc2019}. We measure $\xi$ on halos from the following versions of TNG300: (i) hydrodynamic simulation, (ii) the DMO simulation, (iii) the DMO simulation with the global ``abundance matching" halo mass correction, (iv) and the DMO simulation with the environment-dependent halo mass correction. In this figure, the corrections are done on a halo-by-halo basis; in other words, we do not use our polynomial fits to make these corrections, but rather we directly assign DMO halos the exact masses of the corresponding ``abundance matched" hydrodynamic halos. In this way, we can assess the ability of a given mass correction to reproduce the bias seen in the full-physics version of TNG300 without introducing any error due to our fits. In the bottom panel of Figure~\ref{fig:xi} we plot the ratio of each subsequet $\xi$ (DMO, DMO with ``abundance matching" correction, or DMO with environment dependent correction) to the hydrodynamic $\xi$.

For both mass samples, the DMO halos do not exhibit the same correlation function as do the hydrodynamic halos; the residual panel shows that the difference in $\xi$ for the low (high) mass sample is about 8\% (7\%) at the smallest scales, and for both mass samples it is about 4\% for all scales above $2 h^{-1}\mathrm{Mpc}$. After applying the global ``abundance matching” halo mass correction, the difference in $\xi$ reduces to about 2\% for both mass samples for all scales above $2 h^{-1} \mathrm{Mpc}$ (the difference is slightly larger for the high mass sample). This difference reaches about 6\% on the smallest scales for both mass samples. We emphasize that this lingering discrepancy in $\xi$ after applying the abundance matching mass correction is not due to the accuracy of our seventh-order polynomial fit, because in this figure we are not using our fits, but rather matching individual halos. In other words, at this point the global halo mass function has been perfectly corrected, so the remaining differences in $\xi$ must be due to other factors.

After applying the environment-dependent halo mass correction, the discrepancies in $\xi$ shrink once again. 
For both mass samples, the discrepancy between the environment-dependent corrected DMO halo correlation function and the hydrodynamic halo correlation function is less than 1\% for all scales above $2 h^{-1}\mathrm{Mpc}$. At smaller scales, the environment-dependent correction is still an improvement over the original correction or no correction, although the discrepancy still reaches about 5 or 6\% at the smallest scales we consider. This is due to the fact that halo exclusion occurs at smaller scales in the hydrodynamic simulation than it does in the DMO simulation. Because our halo mass corrections have no impact on the sizes or positions of halos, they cannot improve the halo correlation function until it reaches a scale where both the DMO and hydrodynamic simulations are not lacking halo pairs due to halo exclusion.

These results indicate that the impact of baryonic physics on the halo mass function is dependent on the environment of the halo, which in turn affects the ability of a DMO simulation to reproduce the halo clustering observed in hydrodynamic simulations. Our original ``abundance matching" halo mass corrections are able to reproduce the global halo mass function seen in hydrodynamic simulations, and are able to reproduce the halo clustering to within a few percent. However, if higher accuracy is needed, it is important to account for environment-dependent baryonic effects. Our environment-dependent halo mass corrections reproduce both the global and the conditional halo mass function from hydrodynamic simulations, and they reproduce the halo clustering to within less than 1\%.

In Table~\ref{tab:envfits0} we provide our environment-dependent halo mass corrections for $M_{200b}$ and $M_\mathrm{vir}$ halos in Illustris, TNG, and EAGLE at $z=0$. Additionally, corrections for the remaining halo definitions and redshifts are available in the machine-readable format, as well as in our \textsc{python} module (\url{https://github.com/gbeltzmo/halo_mass_correction}). 

\begin{figure*}
\centering
	\includegraphics[width=\textwidth]{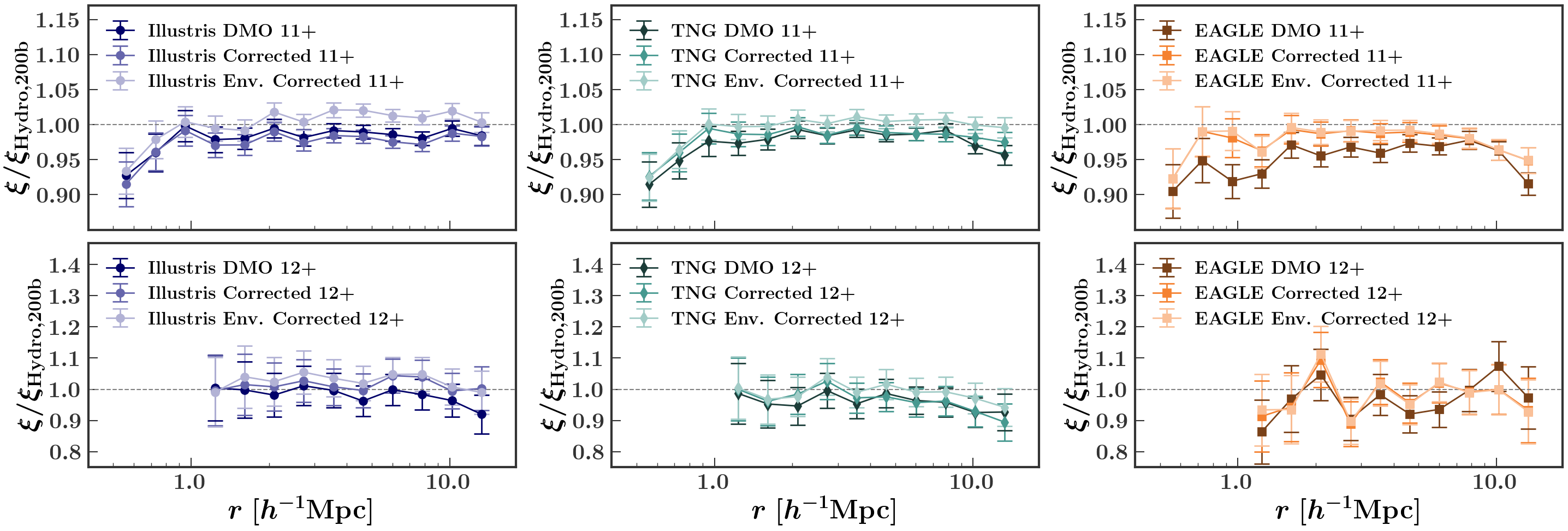}
    \caption{The discrepancy in halo correlation functions for $M_{200b}$ halos at $z=0$ in Illustris-1 (left), TNG100-1 (middle), and EAGLE (right). In each panel, we show the discrepancy between the hydrodynamic halo correlation function and that in DMO (dark), DMO with the original halo mass correction (medium), and DMO with the environment dependent halo mass correction (light). The top panels are the correlation function discrepancies for halos greater than $10^{11} h^{-1} M_\odot$, and the bottom panels are for halos greater than $10^{12} h^{-1} M_\odot$.}
    \label{fig:clustering_correction}
\end{figure*}

Figure~\ref{fig:clustering_correction} shows the results of applying these environment-dependent halo mass corrections to Illustris (left), TNG (middle), and EAGLE (right). Each panel shows the discrepancy between the hydrodynamic correlation function and that in DMO (dark), DMO with the original halo mass correction (medium), and DMO with the environment dependent halo mass correction (light). In this figure, the mass corrections used are the fits given in Tables~\ref{tab:fits0} and \ref{tab:envfits0} for $M_{200b}$ halos at $z=0$. The top panels show the correlation function discrepancies for halos above than $10^{11} M_\odot$, and the bottom panels show the same for halos above $10^{12} M_\odot$. (We do not show halos below $10^{11} M_\odot$ because in all three simulations the environment dependent corrections do not deviate from the original corrections below $10^{11} M_\odot$. Additionally, the clustering of low mass halos is more complicated, and its investigation is beyond the scope of this paper.)

In each simulation, the DMO halos exhibit some clustering discrepancy compared to the hydrodynamic halos. For the lower mass sample (top panels), DMO underestimates the clustering by about 3\% on average in Illustris, and TNG, and by about 5\% on average in EAGLE. In Illustris, the original mass correction actually makes the clustering slightly worse, while the environment-dependent correction is an improvement on small-scales, and a slight overcorrection on large scales. In TNG, the original mass correction provides little to no improvement over DMO, but the environment-dependent correction improves the clustering almost completely. In EAGLE, the original and the environment-dependent corrections are in very close agreement, and both improve the clustering almost entirely compared to DMO.

For the higher mass sample (bottom panels), the error bars are large due to a lack of high mass halos. However, in each case, the DMO simulations underestimate the clustering of halos compared to the hydrodynamic simulations. This discrepancy is smallest in Illustris, and as a result, both the original and the environment-dependent mass corrections yield a slight overestimation of clustering. In TNG, the original mass correction provides little to no improvement, while the environment-dependent correction once again improves the clustering almost completely. In EAGLE, both the original and the environment-dependent corrections provide a slight improvement compared to DMO.

Applying these corrections to a large DMO simulation is slightly more complicated, because it requires calculating the large-scale environment for each halo of interest (only those above $10^{10} h^{-1} M_\odot$), and then separating these halos into ``high" and ``low" density environments based on the median environment $\delta_\mathrm{med}$, and applying the corresponding mass correction. (In our module, an environment argument must be passed, which can be ``all," ``high," or ``low," wherein the code assumes it is applying the correction to all halos, only halos in high density environments, or only halos in low density environments, respectively.) 

We have provided the $\delta_\mathrm{med}$ that we found for each simulation and halo definition in Table~\ref{tab:envfits0}. If applying our environment dependent corrections, we suggest using the $\delta_\mathrm{med}$ that we provide for a given correction, rather than finding a new $\delta_\mathrm{med}$ that is specific to your sample. This is because our corrections are based on splitting halos into high- and low-density environments using a $\delta_\mathrm{med}$ for all halos above $10^{10} h^{-1} M_\odot$. However, $\delta_\mathrm{med}$ does increase with halo mass. Thus, for a halo catalogue that only includes halos with masses above $10^{12} h^{-1} M_\odot$, for example, the pertinent $\delta_\mathrm{med}$ would be higher. This means that halos which we identified as being in high-density environments might now be identified as being in low-density environments. In this case, it would not be appropriate to apply our environment-dependent halo mass corrections to this sample. This is admittedly a limitation of our environment-dependent halo mass corrections. The effect of using the ``wrong" value of $\delta_\mathrm{med}$ is very slight, but we still recommend using the value of $\delta_\mathrm{med}$ that we provide to ensure that the mass corrections are applied to the appropriate halos.

We once again provide upper mass limits for our halo mass corrections, and emphasize the importance of not applying the corrections to any halos above these mass limits or below $10^{10} h^{-1} M_\odot$.

\section{Halo Mass - Concentration Relation} \label{sec:concentration}
Halo modeling often uses the internal structure of halos to determine the placement of satellite galaxies within the halo \citep[e.g.][]{Zehavi2011, Zentner2014}. This often involves assuming that the spatial distribution of satellite galaxies within halos follows an NFW profile \citep{Navarro1996}, which includes a parameter for the concentration of the halo. Up to this point, we have investigated and quantified the impact of baryonic physics on the halo mass function, as well as the halo correlation function. We now examine the impact of baryonic physics on halo concentration.

The concentration $c$ of a halo is defined as the ratio of the virial radius $R_\mathrm{vir}$ of the halo to the scale radius $R_s$ \citep{Navarro1997}. The relationship between halo mass and concentration has been previously studied in simulations, and it has been found that halo concentration has a weak power-law dependence on halo mass, with a slope of approximately $-0.13$ \citep{Bullock2001}. This relationship has a great deal of scatter, and is predicated on the assumption that the density within a dark matter halo follows an NFW profile.

\citet{Ragagnin2019} investigated the dependence of halo concentration on mass and redshift in the Magneticum hydrodynamic simulations, and later \citet{Ragagnin2021} examined the cosmology dependence of halo masses and concentrations in hydrodynamic simulations. \citet{Bose2020} examined the mass and concentration of host halos in IllustrisTNG as they relate to velocity dispersion. Additionally, \citet{Wang2020} studied the density profiles of early-type galaxies (ETGs) in IllustrisTNG at $z=0$, and found that the profiles are steeper in the hydrodynamic simulation than their counterparts in the DMO simulation. They also found that the density profiles of the ETG dark matter halos are well described by steeper than NFW profiles.

In order to investigate the effect of baryonic physics on the halo mass - concentration relation in Illustris, TNG100, and EAGLE, we must first calculate the concentration for each halo (above $10^{10} h^{-1} M_\odot$) in these simulations. While concentration is not directly provided in any of these simulations' halo catalogues, we can estimate the concentration from the various halo mass definitions that are provided: $M_{200b}$, $M_{200c}$, and $M_{500c}$. The method works as follows. Assuming that each halo obeys an NFW profile, we can integrate the profile and use the virial mass and virial radius of the halo to determine the scale density $\rho_0$ of the halo, substituting $R_\mathrm{vir}/c$ for $R_s$:\\

\begin{figure*}
\centering
	\includegraphics[width=\textwidth]{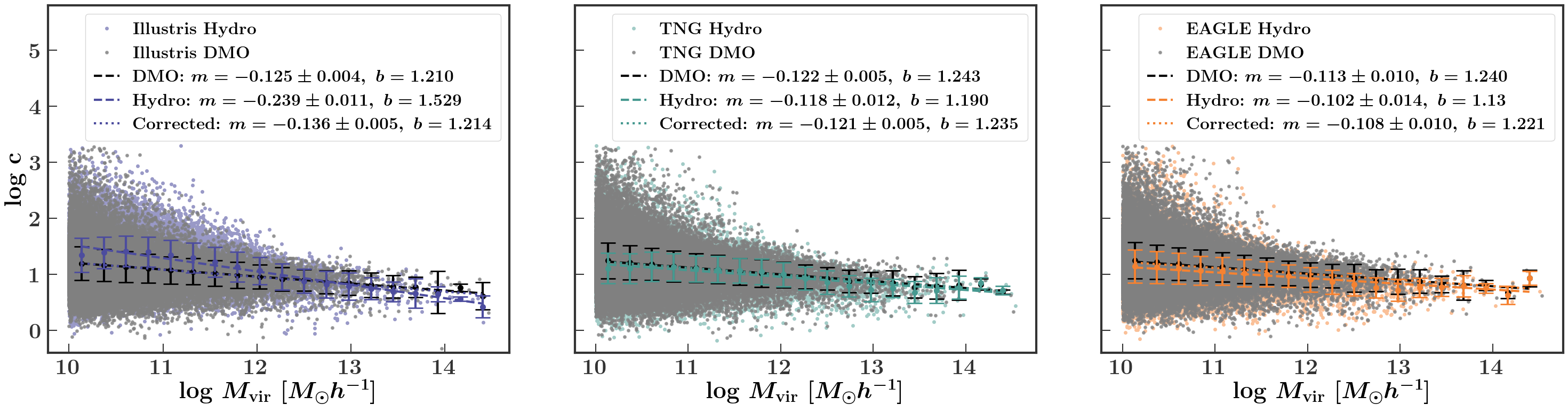}
    \caption{Concentration as a function of $M_\mathrm{vir}$ halo mass (in units of $h^{-1} M_\odot$) for Illustris (left), TNG100 (center), and EAGLE (right) halos at $z=0$. DMO halos are plotted in gray for each simulation, while hydrodynamic halos are plotted in blue (Illustris), green (TNG100), and orange (EAGLE). The larger points in each panel are the mean concentrations in bins of halo mass for each DMO (black) and hydrodynamic (blue/green/orange) simulation, along with their standard deviations. Additionally, we fit a line to these binned points for each simulation, with the corresponding slope and y-intercept shown in the legend of each panel.}
    \label{fig:concentration}
\end{figure*}

$M_\mathrm{vir} = \int_{0}^{R_\mathrm{vir}} 4 \pi r^2 \rho (r) dr$\\

$\rho_0 = \frac{M_\mathrm{vir} c^3}{4\pi R_\mathrm{vir}^3 [ln(1+c) - \frac{c}{1+c}]}$. \\

We can then substitute this to solve for any other halo mass $M_h$ and radius $R_h$ (e.g. $M_{200b}$ and $R_{200b}$): \\

$M_h = \frac{M_\mathrm{vir}}{ln(1+C) - \frac{c}{1+c}} [ln(1 + cR_h/R_\mathrm{vir}) - \frac{cR_h}{cR_h + R_\mathrm{vir}}]$. \\

We can rearrange this equation to find the ratio of $M_\mathrm{vir}$ to $M_h$ as a function of $R_\mathrm{vir}$, $R_h$, and $c$: \\

$ \frac{M_\mathrm{vir}}{M_h} = \frac{ln(1+c) - \frac{c}{1+c}} {ln(1 + cR_h/R_\mathrm{vir}) - \frac{cR_h}{cR_h + R_\mathrm{vir}}}$. \\

For each of our simulations, we have three known values of $M_h$: $M_{200b}$, $M_{200c}$, and $M_{500c}$, and three corresponding values of $R_h$. To find the value of concentration for each halo, we loop over possible values of $c$ between 0 and 2000 and find the one that minimizes the sum of the squared fractional difference between the left- and right-hand side of the previous equation over the three halo definitions. In other words, \\ 

$A = \frac{M_\mathrm{vir}}{M_h}$ and $B = \frac{ln(1+c) - \frac{c}{1+c}} {ln(1 + cR_h/R_\mathrm{vir}) - \frac{cR_h}{cR_h + R_\mathrm{vir}}}$, \\

and we find the value of $c$ for each halo that minimizes \\

$ (\frac{B_{200b} - A_{200b}}{A_{200b}})^2 + (\frac{B_{200c} - A_{200c}}{A_{200c}})^2 + (\frac{B_{500c} - A_{500c}}{A_{500c}})^2 $. \\

We have tested this method of determining concentration on a DMO halo catalogue for which we know the concentration values for each halo (calculated by the \textsc{rockstar}; \citealt{Behroozi2013}). For this particular DMO halo catalogue, using the given values of concentration led to a halo mass - concentration relation with a slope of $-0.104 \pm 0.004$, while using our method of determining concentration led to a slope of $-0.098 \pm 0.006$. Therefore, we can conclude that our method leads to the correct overall halo mass - concentration relation.

After confirming its accuracy, we applied this method to each of our hydrodynamic and DMO simulations to determine the concentration of each halo. In Figure~\ref{fig:concentration} we plot log(c) as a function of log($M_\mathrm{vir}$) (in units of $h^{-1} M_\odot$) for Illustris (left), TNG100 (middle), and EAGLE (right) halos. The results for the DMO simulations are plotted as gray points, while the results for the hydrodynamic simulations are plotted in blue, green, and orange, respectively. We then bin the points by mass, and plot the mean and standard deviation of these bins (in black for DMO or blue/green/orange for the hydrodynamic simulations). We subsequently fit a line to these means, which we plot with a dashed line. (Note that the dashed line is not connecting the points, but rather is a fit to the points.) The slope and y-intercept for each simulation are given in Figure~\ref{fig:concentration}, along with the standard error in the slope. 

Based on these results, we can see that all three DMO simulations have halo mass - concentration relations that are consistent with each other and with what we expect from previous studies. The Illustris DMO simulation has a halo mass - concentration relation with a slope of $-0.125 \pm 0.004$, while the TNG DMO simulation has a halo mass - concentration relation with a slope of $-0.122 \pm 0.005$, and the EAGLE DMO simulation has a halo mass - concentration relation with a slope of $-0.113 \pm 0.010$ (only slightly shallower than expected). 

The Illustris hydrodynamic simulation has a halo mass - concentration relation with a slope of $-0.239 \pm 0.011$, which is steeper than we would expect, and not consistent with that of the DMO simulation. The TNG100 hydrodynamic simulation has a halo mass - concentration relation with a slope of $-0.118 \pm 0.012$, and the EAGLE hydrodynamic simulation has a slope of $-0.102 \pm 0.014$, both of which are consistent with their corresponding DMO simulations. In each of these cases, the scatter among both the DMO and hydrodynamic halos is quite large, but there does not appear to be a systematic offset between the DMO and hydrodynamic distributions for any simulation. Based on these results, we can conclude that in TNG and EAGLE, baryonic physics does not significantly impact the halo mass - concentration relation, while in Illustris, baryonic physics results in a slightly steeper halo mass - concentration relation compared to that in Illustris-Dark.

We would also like to know what the halo mass - concentration relation would look like in each of these simulations if we corrected the DMO halo masses, but did not alter their concentrations. Another way of saying this is if someone were to apply our mass correction to a large DMO box, but not change the halo concentration parameter, how would that impact the halo mass - concentration relation? To investigate this, we apply our ``abundance matching" halo mass correction to our DMO halos for each simulation, and then plot the original DMO halo concentrations as a function of these corrected masses. In this case, we once again do a direct halo-by-halo correction rather than applying our seventh-order polynomial fit, in order to perfectly reproduce the mass function from the hydrodynamic simulation. Thus, the concentration (which we found via the method outlined above using the original halo masses and radii) of the most massive DMO halo is plotted against the mass of the most massive hydrodynamic halo (i.e. the ``corrected" DMO halo mass), and so on. We then find the mean concentration in bins of mass, and fit a new line to this relationship. This new fit is plotted as a dotted line in each panel of Figure~\ref{fig:concentration}, and the slope and y-intercept are given in the legend of each panel. 

In each case, the ``corrected" halo mass - concentration relation has a slope that is still consistent with what we would expect for halos obeying an NFW profile. In all three cases, applying the correction achieves a slope that is slightly closer to the hydrodynamic results. This is most significant in Illustris, where the slope becomes steeper, although still does not agree fully with the hydrodynamic results. In TNG and EAGLE, the DMO, hydrodynamic, and corrected results are all consistent with one another.

Based on this, for any work relying on halo masses and concentrations from a DMO simulation, we believe that depending on how one wants to employ the halo mass - concentration relation, one can either correct only the halo masses (and leave the concentrations untouched, with little impact on the results), or one can correct the halo concentrations as well using the slopes from the full-physics versions of Illustris, TNG, or EAGLE (shown in Figure~\ref{fig:concentration}). Either way, the impact of baryonic physics on the concentration-mass relation is likely to be small.

\section{Conclusions} \label{sec:conclusions}
The implementation of baryonic physics in hydrodynamic simulations results in halo mass functions that are generally shifted to lower masses than those produced by dark matter only simulations. This is because stellar and AGN feedback removes baryonic particles (as well as some dark matter particles) from halos over time. This effect varies with halo mass: stellar feedback has more of an impact on lower mass halos, while AGN feedback has more of an impact on higher mass halos. Additionally, particularly efficient star formation can serve to \textit{increase} the masses of some hydrodynamic halos (compared to their DMO counterparts). 

In this work, we have quantified the relationship between the masses of halos in hydrodynamic simulations and those in corresponding DMO simulations. The impact of baryonic physics on the halo mass function depends on redshift, as well as halo definition. Additionally, because different hydrodynamic simulations contain different baryonic physics prescriptions, the halo mass discrepancy between hydrodynamic and DMO halos varies widely from one simulation to the next. Furthermore, the impact of baryonic physics on halo mass depends somewhat on the large-scale environment of the halo: in Illustris and TNG, halos in low-density environments are more impacted by baryonic physics, leading to greater discrepancies in their masses compared to their DMO counterparts. In EAGLE, halos in high-density environments exhibit a greater mass discrepancy between hydrodynamic and DMO simulations. This indicates that in hydrodynamic simulations in general, the strength of feedback (and its ability to impact halo mass) has a dependence on the density of the halo's environment.

We have found that this relationship as a function of DMO halo mass is well-fit with a seventh-order polynomial. We provide these fits for Illustris, IllustrisTNG100, and EAGLE halos for $M_{200b}$, $M_\mathrm{FoF}$, $M_\mathrm{vir}$, $M_{200c}$, and $M_{500c}$ halos at $z=0,1,$ and 2. These fits are based on matching halos by mass (i.e. ``abundance matching") across hydrodynamic and DMO simulations. In other words, these are the corrections one would need to apply to the halos from Illustris-Dark, for example, to reproduce the halo mass function in the full-physics version of Illustris. We also provide these same fits after taking halo environment into account, which can be used to reproduce the conditional as well as the global mass function. Furthermore, we have shown that these corrections for halo mass also reproduce the large-scale clustering of halos, though the environment-dependent corrections are required to achieve an accuracy better than 2\%. Finally, we have shown that baryonic effects do not impact the halo concentration-mass relation substantially. Our halo mass corrections are publicly available as a \textsc{python} module at \url{https://github.com/gbeltzmo/halo_mass_correction}. 

Any work relying on halo catalogues from DMO simulations (e.g., halo occupation distribution modeling, conditional luminosity function modeling, stellar-to-halo mass relation modeling, etc.) could potentially be impacted by inaccuracies in the halo mass function. In particular, as these types of analyses start to be used more to constrain cosmological parameters, it is imperative that any conclusions are robust to changes in the halo mass function on the order of what we find in this work. For example, a DMO simulation created with a given set of cosmological parameters can produce a halo mass function that, after being adjusted with one of our mass corrections, resembles a HMF from a different cosmology. Thus, without understanding the uncertainty in the halo mass function due to baryonic physics, it may be challenging to distinguish between these cosmological models.

We recommend that any future work utilizing a halo catalogue from a DMO simulation repeat their analysis after applying at least one of our halo mass corrections (and ideally more than one). This will provide a rough estimate of the systematic uncertainty in one's results due to baryonic effects on the halo mass function. We make no assumptions about which of the three hydrodynamic simulations used in this work produces the ``correct" halo mass function, but rather provide our halo mass corrections as a method for determining the robustness of one's modelling results to changes in the halo mass function.

\acknowledgments
We acknowledge the Virgo Consortium for making their simulation data available. The EAGLE simulations were performed using the DiRAC-2 facility at Durham, managed by the ICC, and the PRACE facility Curie based in France at TGCC, CEA, Bruy\`eres-le-Ch\^atel. Some of the computational facilities used in this project were provided by the Vanderbilt Advanced Computing Center for Research and Education (ACCRE). This project has been supported by the National Science Foundation (NSF) through Award (AST-1909631).This research has made use of NASA's Astrophysics Data System, as well as \textsc{python} (\url{https://www.python.org/}), the \textsc{ipython} package \citep{PER-GRA:2007}, \textsc{scipy} \citep{jones_scipy_2001}, \textsc{numpy} \citep{van2011numpy}, and \textsc{matplotlib}, a Python library for publication quality graphics \citep{Hunter2007}. These acknowledgements were compiled using the Astronomy Acknowledgement Generator (\url{http://astrofrog.github.io/acknowledgment-generator/}). We especially thank Jolanta Zjupa for her help, and our anonymous referee for constructive comments that helped us improve this paper.

\bibliography{halos}
\bibliographystyle{aasjournal}

\end{document}

%% file: simulations_table.tex
\begin{deluxetable*}{ccccccccccc}
\tablenum{1}
\tablecaption{Simulation parameters\label{tab:simulations}}
\tablewidth{0pt}
\tablehead{
\colhead{Simulation} & \colhead{$L_\mathrm{box}$} & \colhead{$N_\mathrm{DM}$} & \colhead{$m_\mathrm{DM}$} &
\colhead{$m_\mathrm{gas}$} & \colhead{$h$} & \colhead{$\Omega_{m}$} & \colhead{$\Omega_{\Lambda}$} & \colhead{$\Omega_{b}$} & \colhead{$\sigma_8$} & \colhead{$n_s$} \\
\colhead{Name} & \colhead{$(h^{-1} \mathrm{Mpc})$} & \colhead{} & \colhead{$(h^{-1} M_\odot)$} &
\colhead{$(h^{-1} M_\odot)$} & \colhead{} & \colhead{} & \colhead{} & \colhead{} & \colhead{} & \colhead{}
}
\startdata
Illustris-1 & 75 & $1820^3$ & $4.4 \times 10^6$ & $9.2 \times 10^5$ & $0.704$ & $0.2726$ & $0.7274$ & $0.0456$ & $0.809$ & $0.963$ \\
TNG100-1 & 75 & $1820^3$ & $5.1 \times 10^6$ & $9.4 \times 10^5$ & $0.6774$ &  $0.3089$ & $0.6911$ & $0.0486$ & $0.8159$ & $0.9667$ \\
TNG300-1 & 205 & $2500^3$ & $4.0 \times 10^7$ & $7.6 \times 10^6$ & $0.6774$ &  $0.3089$ & $0.6911$ & $0.0486$ & $0.8159$ & $0.9667$ \\
EAGLE & 67.77 & $1504^3$ & $6.6 \times 10^6$ & $1.2 \times 10^6$ & $0.6777$ & $0.307$ & $0.693$ & $0.04825$ & $0.8288$ & $0.9611$
\enddata
\tablecomments{The columns show (from left to right): simulation name, box size, number of dark matter particles, dark matter particle mass (for the hydrodynamical run), gas particle mass, redshift used, and cosmological parameters. The dark matter particle mass for Illustris-1-Dark is $5.3 \times 10^6 (h^{-1} M_\odot)$; for TNG100-1-Dark it is $6.0 \times 10^6 (h^{-1} M_\odot)$; for TNG300-1-Dark it is $7.0 \times 10^7 (h^{-1} M_\odot)$; and for EAGLE Dark it is $7.5 \times 10^6 (h^{-1} M_\odot)$.}
\end{deluxetable*}